\documentclass[journal,12pt,draftclsnofoot,onecolumn]{IEEEtran}
\IEEEoverridecommandlockouts
\ifCLASSINFOpdf
\else
\fi
\usepackage{amsmath,graphicx,cite,amssymb,bm,amsthm,booktabs,lipsum}
\newtheorem{lem}{Lemma}
\newtheorem{theorem}{Theorem}

\begin{document}
\begin{titlepage}
\begin{center}
\vspace*{-2\baselineskip}
\begin{minipage}[l]{7cm}
\flushleft
\includegraphics[width=2 in]{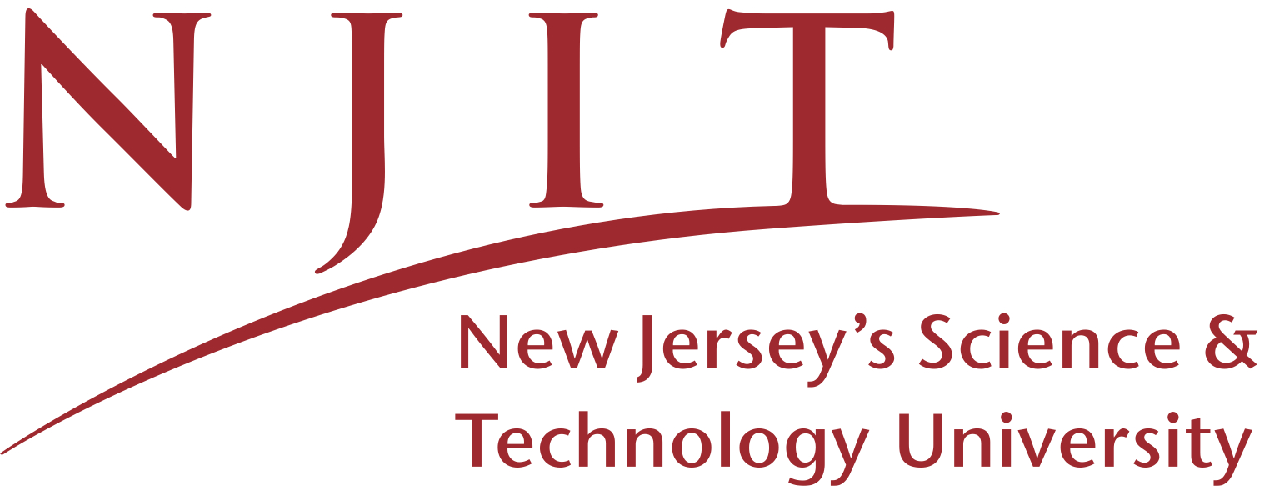}
\end{minipage}
\hfill
\begin{minipage}[r]{7cm}
\flushright
\includegraphics[width=1 in]{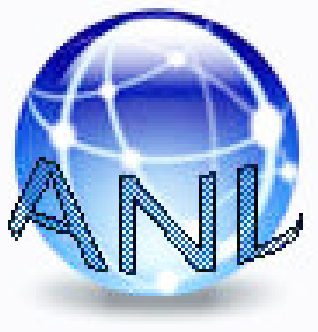}
\end{minipage}

\vfill

\textsc{\LARGE PRIMAL: PRofIt Maximization Avatar pLacement for Mobile Edge Computing\\[12pt]}
\vfill
\textsc{
\LARGE  XIANG SUN \\ NIRWAN ANSARI}\\
\vfill
\textsc{\LARGE TR-ANL-2015-007\\[12pt]
\LARGE Oct 14, 2015}\\[1.5cm]
\vfill
{ADVANCED NETWORKING LABORATORY\\
 DEPARTMENT OF ELECTRICAL AND COMPUTER ENGINEERING\\
 NEW JERSY INSTITUTE OF TECHNOLOGY}
\end{center}
\end{titlepage}

\title{PRIMAL: PRofIt Maximization Avatar pLacement for Mobile Edge Computing}
\author{\IEEEauthorblockN{Xiang Sun,~\IEEEmembership{Student Member,~IEEE}, and Nirwan Ansari,~\IEEEmembership{Fellow,~IEEE}}
}
\maketitle

\begin{abstract}
We propose a cloudlet network architecture to bring the computing resources from the centralized cloud to the edge. Thus, each User Equipment (\emph{UE}) can communicate with its Avatar, a software clone located in a cloudlet, with lower end-to-end (\emph{E2E}) delay. However, UEs are moving over time, and so the low E2E delay may not be maintained if UEs' Avatars stay in their original cloudlets. Thus, live Avatar migration (i.e., migrating a UE's Avatar to a suitable cloudlet based on the UE's location) is enabled to maintain low E2E delay between each UE and its Avatar. On the other hand, the migration itself incurs extra overheads in terms of resources of the Avatar, which compromise the performance of applications running in the Avatar. By considering the gain (i.e., the E2E delay reduction) and the cost (i.e., the migration overheads) of the live Avatar migration, we propose a PRofIt Maximization Avatar pLacement (\emph{PRIMAL}) strategy for the cloudlet network in order to optimize the tradeoff between the migration gain and the migration cost by selectively migrating the Avatars to their optimal locations. Simulation results demonstrate that as compared to the other two strategies (i.e., Follow Me Avatar and Static), PRIMAL maximizes the profit in terms of maintaining the low average E2E delay between UEs and their Avatars and minimizing the migration cost simultaneously.
\end{abstract}

\IEEEpeerreviewmaketitle

\section{Introduction}
Recent mobile applications, such as augmented reality, image processing and speech recognition, become resource intensive and drain User Equipments' (\emph{UE}s') batteries very quickly. Mobile Cloud Computing (\emph{MCC}) \cite{1},\cite{2} has been proposed to offload applications' workloads from UEs to the cloud in order to not only reduce energy consumption of UEs but also accelerate the execution time of the applications. MCC reduces the UE's computational cost at the expense communications cost, i.e., the UE frequently interacts with the cloud by offloading its application workloads. Thus, it is not efficient to do the application offloading if the End-to-End (\emph{E2E}) delay between a UE and the cloud is unbearable. The cloudlet architecture is introduced to reduce the E2E delay, i.e., the computing resources are moved from the remote cloud to the local cloudlet, which is a tiny version of a data center residing close to UEs, so that UEs can access the computing resources via the Local Area Networks (\emph{LANs}) with lower E2E delay \cite{2}.

\begin{figure}[!htb]
	\centering	
	\includegraphics[width=1.1\columnwidth]{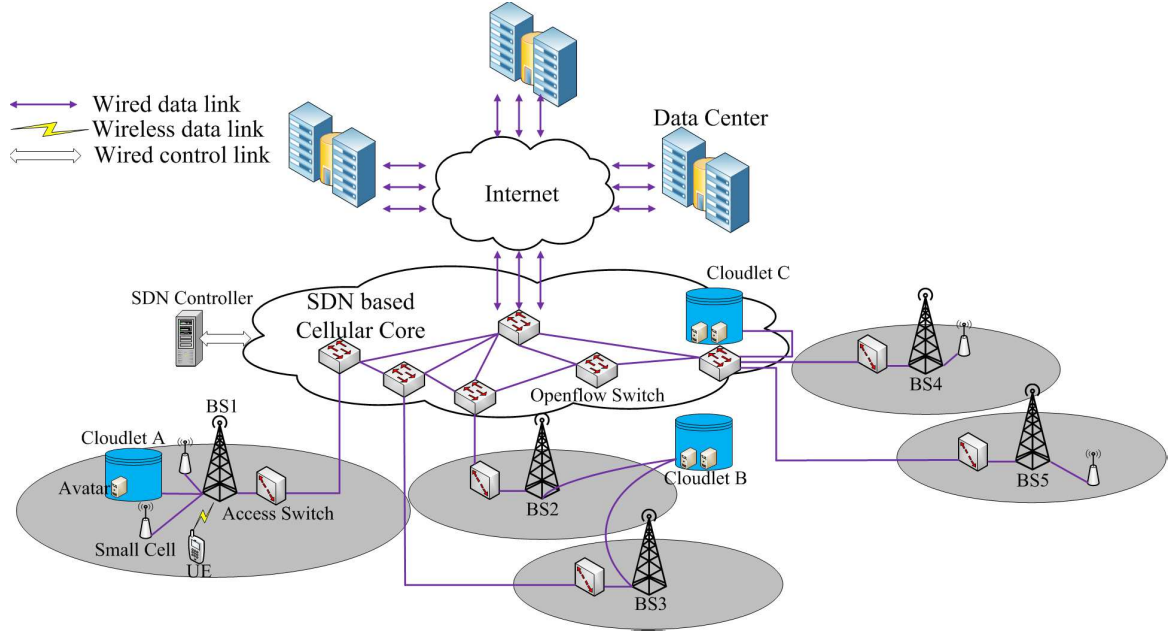}
	\caption{The cloudlet network architecture.}	
	\label{fig1}
\end{figure}

To reap benefits of the cloudlet, we propose the cloudlet network architecture, as shown in Fig. \ref{fig1}, in order to provide ubiquitous computing resources to UEs and at the same time maintain low E2E delay. Since the existing LTE network infrastructure can provide seamless connection between a UE and a base station (\emph{BS}), each BS is connected to a cloudlet via high speed fibers so that UEs can utilize computing resources in the cloudlets with one wireless hop delay. Moreover, each UE subscribes one Avatar, a high performance Virtual Machine (\emph{VM}) in the cloudlet, which provides extra computing resources and storage space. Avatars are software clones of their UEs and always available to UEs when UEs are moving from one coverage area to another \cite{3}. Assigning a specific Avatar to each UE in the cloudlet provides hardware isolation by securely running each UE's application workloads on a shared physical hardware. On the top of the cloudlets, Software Defined Network (\emph{SDN}) based cellular core network has been proposed in the cloudlet network architecture to provide efficient and flexible communications paths between Avatars in different cloudlets as well as between UEs in different BSs \cite{4},\cite{5}. Moreover, every UE and its Avatar in the cloudlet can communicate with public data centers (e.g., Amazon EC2) and Storage Area Networks (\emph{SAN}s) via the Internet in order to provision scalability, i.e., if cloudlets are not available for UEs because of the capacity limitation, UEs' Avatars can be migrated to the remote data centers to continue serving their UEs.

The locations of cloudlets depend on the UE density, i.e., the BS in the hotspot area can own a specific cloudlet (e.g., BS\emph{1} connects to a specific cloudlet, say, Cloudlet \emph{A} in Fig. \ref{fig1}), or the BSs in the rural or suburban area can share the same cloudlet. Thus, the cloudlet can be placed among the BSs (e.g., cloudlet \emph{B} is deployed between BS\emph{2} and BS\emph{3} in Fig. \ref{fig1}) or the cloudlet is directly connected to the switch at the edge of the SDN based cellular core (e.g., cloudlet \emph{C} connects to the edge switch so that BS\emph{4} and BS\emph{5} can share the computing and storage resources of cloudlet \emph{C}).

The cloudlet network architecture not only helps UEs offload their application workloads to their Avatars with lower latency but also facilitates real time big mobile data analysis. Smart UEs, embedded with a rich set of sensors, become a data stream generator producing their users' information (e.g., users' locations, activities, mood and their health information) over time. Analyzing these massive amount of mobile data is not only extremely valuable for market applications, but also potentially benefits the society as a whole \cite{6}. Traditionally, these big mobile data are analyzed within a data center \cite{6.1} by utilizing the distributed computing framework, such as MapReduce \cite{7}, Dryad \cite{8} and Storm \cite{9}. However, transmitting the big mobile data from UEs to the data center through the Internet suffers from the long latency and increases the traffic load of the network. Meanwhile, most of the mobile data are time-sensitive, i.e., the potential value of the mobile data is decreasing as time passes by. Thus, rather than bringing the mobile data to the computing resources, the cloudlet network architecture is proposed to bring the computing resources to the mobile data. In other words, each Avatar locally collects, filters, classifies or even analyzes the raw data stream of its user so that the volume of the mobile data, which need to be transmitted to the remote data center for further analysis, can be reduced substantially or eliminated and the communications latency can be reduced as well. A typical example by utilizing the cloudlet network architecture to analyze the big mobile data is the terrorist localization application, i.e., each Avatar receives the terrorists' photos and runs the face matching algorithm locally to compare the recent photos and videos captured by its user. If matched, the information of the photos/videos, i.e., the locations and timestamps of the photos/videos, would be uploaded to the central server for further processing.

The rest of the paper is organized as follows. In Section II, we propose the live Avatar migration among cloudlets to maintain the low E2E delay between UEs and their Avatars when UEs are moving over time. We design the live Avatar migration gain and cost models to calculate the gain and cost of each Avatar migration, respectively. In Section III, we formulate the novel Avatar placement strategy, referred to as PRofIt Maximization Avatar pLacement (\emph{PRIMAL}), to maximize the migration profit in terms of optimizing the tradeoff between the migration gain and the migration cost. In Section IV, we demonstrate the performance of the proposed strategy. The conclusion is presented in Section V.

\section{Live Avatar Migration}
UEs frequently communicate with their Avatars by transmitting their mobile data (i.e., the application workloads and users' data streams) over time. Thus, deploying the Avatar close to its UE will essentially reduce the E2E delay between the UE and its Avatar, and helps meet the QoS of MCC applications and facilitates big mobile data analysis. However, UEs are moving over time, and so UEs may be far away from their Avatars if the locations of the Avatars are static, i.e., the E2E delay between a UE and its Avatar in the cloudlet might be worse than the E2E delay between a UE and a VM in a remote data center. Therefore, it is necessary to optimize each Avatar's placement based on its UE's location. By taking advantages of the live VM migration in a data center, Avatars can also be migrated among cloudlets over the SDN based celluar core network to alter their locations, and thus the E2E delay between a UE and its Avatar can be reduced by optimizing the Avatars' locations. Nevertheless, Avatar migrations are expensive operations because they incur additional overheads \cite{10,11,12}, i.e., Avatar migrations consume extra resources (e.g., CPU, memory, network, disk I/O resources), which affect the performance of applications running in the Avatars. While migrating the Avatar close to its UE potentially improves the E2E delay, it may introduce humongous migration overheads. In order to measure the profit of the migration, it is important to consider the gain and the cost of the Avatar migration simultaneously.

\subsection{Live Avatar Migration Gain Model}
Different from the traditional live VM migration in a data center (that tries to maximize the resource utilization and reduce the energy consumption of the data center), the benefit of the live Avatar migration is to reduce the E2E delay between a UE and its Avatar, which comprises three parts: first, $T^{access}$, i.e., the E2E delay between a UE and its BS (to which the UE is associated with); second, $T^{core}$, i.e., the E2E delay between the UE's BS and the UE's cloudlet (in which the UE's Avatar is located); third, $T^{cloudlet}$, i.e., the E2E delay within the cloudlet. Since changing the placement of UEs' Avatars does not significantly affect the values of $T^{access}$ and $T^{cloudlet}$, we consider the gain of live Avatar migration as the reduction of $T^{core}$, which is the most important parameter affecting the E2E delay between a UE and its Avatar. In other words, if a UE's Avatar is migrated to the cloudlet which has lower E2E delay to the UE's BS, then the gain of the live Avatar migration is defined as the reduction of $T^{core}$. 

Denote $\bm{\mathcal{I}}$ as the set of UEs/Avatars (note that one UE is associated with one specific Avatar, and thus we equate the set of UEs to the set of Avatars), and $i$ is used to index the UEs and their corresponding Avatars. Denote $\bm{\mathcal{J}}$ and $\bm{\mathcal{K}}$ as the set of cloudlets and BSs in the network, respectively, and $j$ and $k$ are used to index the cloudlets and BSs, respectively. Denote $x_{i,j}$ as the binary variable to indicate whether Avatar $i$ is located in cloudlet $j$ (i.e., $x_{i,j}=1$) or not. Meanwhile, $y_{i,k}$ is used to indicate whether UE $i$ is associated with BS $k$ (i.e., $y_{i,k}=1$) or not. By taking the advantage of the SDN network, the E2E between the $j$th ($j \in {\bm{\mathcal{J}}}$) cloudlet and the $k$th ($k \in {\bm{\mathcal{K}}}$) BS in the cloudlet network, denoted as $d_{j,k}$, can be measured by the SDN controller in each time slot \cite{13},\cite{13.1}. Thus, the E2E delay between UE $i$'s BS and UE $i$'s cloudlet can be derived as follows:
		\begin{equation} \small
		T_i^{core} = \sum\limits_{j = 1}^{\left| {\bm{\mathcal{J}}} \right|} {\sum\limits_{k = 1}^{\left| {\bm{\mathcal{K}}} \right|} {{x_{i,j}}{y_{i,k}}{d_{j,k}}}}.
		\label{eq1}
		\end{equation}

Assuming that reducing one unit of the E2E delay increases one unit of the gain for a UE, then given UE $i$'s location (denoted as $y_{i,k}^{t+1}$) in the next time slot, the migration gain model is defined as the amount of E2E delay reduction achieved by the case that UE $i$'s Avatar migrates to the location of $x_{i,j}^{t+1}$ in the next time slot as compared to the case that UE $i$'s Avatar stays in the current location (i.e., $x_{i,j}^{t}$) in the next time slot, i.e.,
		\begin{equation} \small
		{r_i} = \sum\limits_{j = 1}^{\left| {\bm{\mathcal{J}}} \right|} {\sum\limits_{k = 1}^{\left| {\bm{\mathcal{K}}} \right|} {\left( {x_{i,j}^t - x_{i,j}^{t + 1}} \right)y_{i,k}^{t + 1}{d_{j,k}}}}.
		\label{eq2}
		\end{equation}

\subsection{Live Avatar Migration Cost Model}
As mentioned previously, Avatar migration may lead to performance degradation of applications running in the Avatar, i.e., although the migration can facilitate the communications between a UE and its Avatar, the available resources for running the Avatar's applications become less, thus compromising the QoS for using the Avatar. Suppose the migration overheads are fixed during the migration process (i.e., the migration consumes the same amount of extra resources in each time slot during the process); if the migration consumes less time, it generates fewer overheads to affect the applications running in the Avatar, i.e., the migration cost is proportional to the total migration time. We set up the live Avatar migration cost model as follows:
\begin{equation} \small
{c_i} = {\kappa _i}T_i^{mig}\sum\limits_{j = 1}^{\left| {\bm{\mathcal{J}}} \right|} {\frac{1}{2}{{\left( {x_{i,j}^t - x_{i,j}^{t + 1}} \right)}^2}},
\label{eq2.1}
\end{equation}
where $c_i$ is the migration cost of Avatar $i$, $T_i^{mig}$ is the total migration time of Avatar $i$, $\kappa _i$ is the cost coefficient that maps the migration time to the cost, and $\sum\limits_{j = 1}^{\left| {\bm{\mathcal{J}}} \right|} {\frac{1}{2}{{\left( {x_{i,j}^t - x_{i,j}^{t + 1}} \right)}^2}}$ indicates whether Avatar $i$ is migrated to another cloudlet (i.e., the summation equals to 1) or not (i.e., the summation equals to 0).

\subsubsection{Total Migration Time}
\begin{figure}[!htb]
	\centering	
	\includegraphics[width=1.0\columnwidth]{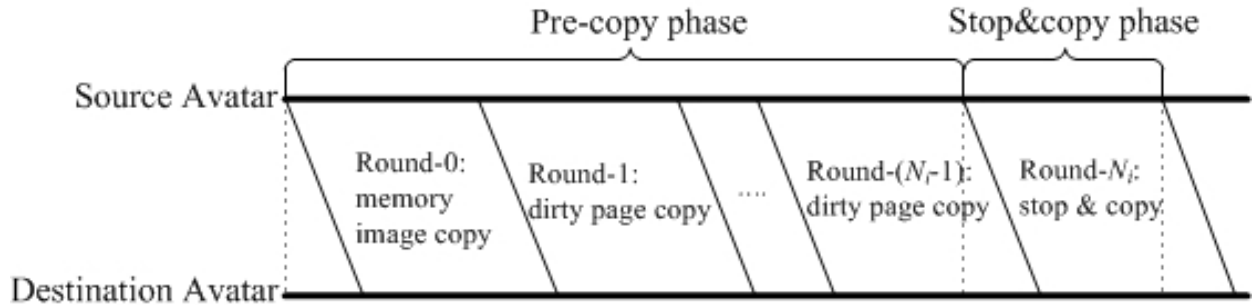}
	\caption{The pre-copy live migration procedure.}	
	\label{fig2}
\end{figure}
The total migration time is different by applying different migration techniques. As an example, we apply the pre-copy live migration technique \cite{14},\cite{15} for migrating Avatars among cloudlets. As shown in Fig. \ref{fig2}., there are two phases during pre-copy live migration, i.e., the pre-copy phase and stop$\&$copy phase \cite{15},\cite{16}. In the pre-copy phase, the whole memory of the source Avatar is transmitted to the destination Avatar in the initial round. For the rest of each round, source Avatar sends the dirty pages, which are generated from the previous round to the destination. Until the number of generated dirty pages is no larger than a predefined threshold, the migration proceeds to the stop-and-copy phase, i.e., the source Avatar stops serving its UE, transmits the rest of the dirty memory pages and informs the destination Avatar to resume services to its UE \cite{12}. 
\begin{lem}
If the bandwidth provisioning for doing migration is constant, given the amount of Avatar $i$'s memory (denoted as $M_i$) and Avatar $i$'s average memory page dirtying rate (i.e., the average number of dirty memory pages generated in each time slot, denoted as $D_i$) during the migration, the time required for executing Avatar $i$'s migration is:
\begin{equation}\small
 T_i^{mig} = \frac{{{M_i}}}{{R - {D_i}}}\left[ {1 - {{\left( {\frac{{{D_i}}}{R}} \right)}^{\left\lceil {{{\log }_{{D_i}/R}}\left( {\frac{{{M^{th}}}}{{{M_i}}}} \right) + 1} \right\rceil  + 1}}} \right]
 \label{eq3}
 \end{equation}
where $R$ is the bandwidth provisioning in terms of the data rate for doing migration ($R > {D_i}$) and $M^{th}$ is the threshold of dirty pages generated. 
\label{lemma1}
\end{lem} 

\begin{IEEEproof}
Suppose there are $N_i$ number of rounds during the migration process and the SDN network provider can guarantee a constant bandwidth in terms of a fixed bit rate for doing migration, the amount of time consumed in the current round $n$ depends on the amount of dirty memory generated in the previous round $n-1$. Thus, we have \cite{17}:
		\begin{equation} \small
		T_{i,n}^{mig} = \frac{{{D_i}}}{R}T_{i,n - 1}^{mig}{\rm{ =  }}{\left( {\frac{{{D_i}}}{R}} \right)^n}T_{i,0}^{mig}{\rm{,   }}\ 1 \le n \le {N_i},
		\label{eq4}
		\end{equation}
where $T_{i,n}^{mig}$ is the time consumption of round $n$ during Avatar $i$'s migration, and $T_{i,0}^{mig}$ is the time consumption of the initial round (i.e., round 0) during which the whole memory of the source Avatar is transmitted to the destination, i.e., $T_{i,0}^{mig}=\frac{{{M_i}}}{R}$. Thus, the time consumption of round $n$ is:
		\begin{equation} \small
		T_{i,n}^{mig} = {\left( {\frac{{{D_i}}}{R}} \right)^n}\frac{{{M_i}}}{R}{\rm{,  }}\ 0 \le n \le {N_i}.
		\label{eq5}
		\end{equation}

The total time consumption of the migration is the sum of the time consumption in each round, i.e.,
		\begin{equation} \small
		\begin{split}\small
		T_i^{mig} &=\sum\limits_{n = 0}^{{N_i}} {T_{i,n}^{mig}}= \sum\limits_{n = 0}^{{N_i}} {\left\{ {{{\left( {\frac{{{D_i}}}{R}} \right)}^n}\frac{{{M_i}}}{R}} \right\}} \\
	    &= \frac{{{M_i}}}{{R - {D_i}}}\left[ {1 - {{\left( {\frac{{{D_i}}}{R}} \right)}^{{N_i} + 1}}} \right].
		\label{eq6}
		\end{split}
		\end{equation}
		
As mentioned before, once the number of the generated dirty pages are no larger than a predefined threshold (i.e., $M^{th}$) in the previous round, then the source Avatar would stop serving its UE and transmit the rest of the dirty pages to the destination Avatar in the last round, i.e., $T_{i,N_i-1}^{mig}R\leq M^{th}$. Based on Eq. \ref{eq5}, we have $T_{i,N_i-1}^{mig}={\left( {\frac{{{D_i}}}{R}} \right)^{N_i-1}}\frac{{{M_i}}}{R}{\rm}$, and so we can derive that if $R > {D_i}$, ${N_i} \le {\log _{{D_i}/R}}\left( {\frac{{{M^{th}}}}{{{M_i}}}} \right) + 1$ (note that if $R \le D_{i}$, then ${N_i} \to  + \infty$). Since the number of the migration rounds should be an integer value, we have:
\begin{equation} \small
{N_i} = \left\lceil {{{\log }_{{D_i}/R}}\left( {\frac{{{M^{th}}}}{{{M_i}}}} \right) + 1} \right\rceil,\ R > {D_i}.
\label{eq7}
\end{equation}

By substituting Eq. \ref{eq7} into Eq. \ref{eq6}, we have Eq. \ref{eq3} and thus prove Lemma \ref{lemma1}.
\end{IEEEproof}

\subsubsection{Cost Coefficient $\kappa _i$}
The cost coefficient $\kappa_i$ in Eq. \ref{eq3} may also vary among Avatars because even if different Avatars generate the same migration overheads in terms of the same migration time, the performance degradation of applications running in different Avatars are also different. The reason is that the live Avatar migration itself can be considered as an I/O intensive application, and so the Avatar migration would have more negative effect on the performance of applications which have higher I/O footprints as compared to the pure CPU intensive applications \cite{11}, i.e., the Avatars running higher I/O applications have higher cost coefficient than the Avatars running lower I/O applications. Based on the above observation, we model the value of $\kappa_i$ to be proportional to the weighted sum of utilization of different resources \cite{11}:
\begin{equation} \small
{\kappa _i}\!=\!\alpha \left( {{w^{net}}u_i^{net}\!+\!{w^{disk}}u_i^{disk}\!+\!{w^{mem}}u_i^{men}\!+\!{w^{cpu}}u_i^{cpu}} \right),
\label{eq8}
\end{equation}
where $u_i^{net}$, $u_i^{men}$, $u_i^{disk}$ and $u_i^{cpu}$ denote the bandwidth, memory, disk I/O and CPU resource utilization of Avatar $i$, respectively; $w^{net}$, $w^{men}$, $w^{disk}$ and $w^{cpu}$ are the migration impact factor of the bandwidth, memory, disk I/O and CPU resource utilization, respectively, indicating the degree of impact of different resources (note that the values of $w^{net}$, $w^{men}$, $w^{disk}$ and $w^{cpu}$ can be derived through  experiments \cite{11}); $\alpha$ is the penalty coefficient that maps the weighted sum of utilization of resources to the cost coefficient. $\alpha$ is an important parameter in the system. Increasing the value of $\alpha$ would increase the ratio of the migration cost to migration gain, and discourage Avatars from doing live migrations. Consequently, the E2E delay would increase if Avatars are not incentivized to do live migration. Thus, $\alpha$ is a parameter to adjust the tradeoff between the E2E delay and the cost for doing live Avatar migration, and can be chosen via experiments by testing users' QoE for utilizing their Avatars. Also, altering the value of $\alpha$ can adjust the traffic in the SDN cellular core, i.e., increasing the value of $\alpha$ would reduce the traffic generated by the migrations and mitigate the traffic load in the SDN cellular core consequently.
 
\section{PRofIt Maximization Avatar pLacement (\emph{PRIMAL})}
In order to increase the gain by facilitating the communications between a UE and its Avatar, the UE's Avatar can be placed in the cloudlet which has lower E2E delay to the UE's BS. However, changing the placement of the UE's Avatar involves live Avatar migration which would degrade the performance of applications running in the Avatar. Thus, we need to design an optimal Avatar placement strategy to optimize the tradeoff between the migration gain and the migration cost by estimating whether it is worth to do the live Avatar migration or not. Denote $f_i$ as the profit of migrating Avatar $i$, i.e., migration gain minus migration cost (Eq. \ref{eq2} - Eq. \ref{eq2.1}):
\begin{gather} \small
\begin{split}
&{f_i} = {r_i} - {c_i} \\ 
&=\! \sum\limits_{j = 1}^{\left| {\bm{\mathcal{J}}} \right|} {\left\{ {\!- 0.5{\kappa _i}T_i^{mig}{{\left( {x_{i,j}^{t + 1}} \right)}^2}\!+\! \left( {{\kappa _i}T_i^{mig}x_{i,j}^t\!-\!\sum\limits_{k = 1}^{\left| {\bm{\mathcal{K}}} \right|} {y_{i,k}^{t + 1}{d_{j,k}}} } \right)x_{i,j}^{t + 1}} \right\}} \\
&+ \sum\limits_{j = 1}^{\left| {\bm{\mathcal{J}}} \right|} {\left\{ {x_{i,j}^t\sum\limits_{k = 1}^{\left| {\bm{\mathcal{K}}} \right|} {y_{i,k}^{t + 1}{d_{j,k}}}  - 0.5{\kappa _i}T_i^{mig}{{\left( {x_{i,j}^t} \right)}^2}} \right\}}.
\label{eq9}
\end{split}
\end{gather}

We assume that each UE's Avatar is homogeneous, i.e., the hardware configuration of each Avatar is the same, and the capacity of each cloudlet is limited, i.e., each cloudlet can only host a fixed number of Avatars, denoted as $s_j$ ($j \in {\bm{\mathcal{J}}}$). Then, we formulate PRIMAL as follows:
	\begin{align}
	&\mathop {\arg \max }\limits_{x_{i,j}^{t + 1}} \sum\limits_{i = 1}^{\left| {\bm{\mathcal{I}}} \right|} {{f_i}}\\ 
	s.t. \ \ &\forall i \in {\bm{\mathcal{I}}},\ \ \sum\limits_{j = 1}^{\left| {\bm{\mathcal{J}}} \right|} {x_{i,j}^{t + 1} = 1}, \\ 
	& \forall j \in {\bm{\mathcal{J}}},\ \ {\rm{  }}\sum\limits_{i = 1}^{\left| {\bm{\mathcal{I}}} \right|} {x_{i,j}^{t + 1}}  \le {s_j},\\	
	& \forall i \in {\bm{\mathcal{I}}},\forall j \in {\bm{\mathcal{J}}},\ \ {\rm{    }}x_{i,j}^{t + 1} \in \left\{ {0,1} \right\}, 
	\end{align}	
where the objective is to maximize the total profit of Avatar live migrations. The first constraint imposes that every UE's Avatar should be allocated in only one cloudlet. The second constraint means that the total number of Avatars assigned to the cloudlet cannot exceed the cloudlet's capacity.
	\begin{theorem}
	The PRIMAL problem is NP-hard.
	\label {thm1}
	\end{theorem} 
	\begin{IEEEproof}
	The objective function of the problem can be transformed into $\mathop {{\mathop{\rm argmax}\nolimits} }\limits_{x_{i,j}^{t + 1}} \sum\limits_{i = 1}^{\left| {\bm{\mathcal{I}}} \right|} {{g_i}}$, where $ \small {g_i}\! =\! \sum\limits_{j = 1}^{\left| {\bm{\mathcal{J}}} \right|} {\left\{ { - 0.5{\kappa _i}T_i^{mig}{{\left( {x_{i,j}^{t + 1}} \right)}^2}\! +\! \left( {{\kappa _i}T_i^{mig}x_{i,j}^t \!-\! \varepsilon \sum\limits_{k = 1}^{\left| {\bm{\mathcal{K}}} \right|} {y_{i,k}^{t + 1}{d_{j,k}}} } \right)x_{i,j}^{t + 1}} \right\}}$. Suppose the capacity of each cloudlet is one (i.e., $\forall j \in {\bm{\mathcal{J}}},\ {s_j} = 1$) and the number of the Avatars is equal to the number of the cloudlets in the network (i.e.,$\left| {\bm{\mathcal{I}}} \right| = \left| {\bm{\mathcal{J}}} \right|$), the original problem can be reformulated as follows:
	\begin{align}
R1:\ \ \ \ \ \ \ &\mathop {\arg \max }\limits_{x_{i,j}^{t + 1}} \sum\limits_{i = 1}^{\left| {\bm{\mathcal{I}}} \right|} {{g_i}}\\ 
		s.t. \ \ &\forall i \in {\bm{\mathcal{I}}},\ \ \sum\limits_{j = 1}^{\left| {\bm{\mathcal{J}}} \right|} {x_{i,j}^{t + 1} = 1}, \\ 
		& \forall j \in {\bm{\mathcal{J}}},\ \ {\rm{  }}\sum\limits_{i = 1}^{\left| {\bm{\mathcal{I}}} \right|} {x_{i,j}^{t + 1}} = 1,\\	
		& \forall i \in {\bm{\mathcal{I}}},\forall j \in {\bm{\mathcal{J}}},\ \ {\rm{    }}x_{i,j}^{t + 1} \in \left\{ {0,1} \right\}.
		\end{align}	

	The problem of $R1$ is a quadratic assignment problem which is proven to be NP-hard \cite{18}. Thus, the quadratic assignment problem is reducible to the PRIMAL problem, i.e., the PRIMAL problem is NP-hard.	
	\end{IEEEproof}
\begin{lem}
The PRIMAL problem is a concave quadratic optimization problem with binary constraints when $\kappa _i T_i^{mig} > 0$.
\label{lemma2}
\end{lem}	
\begin{IEEEproof}
The lemma can be proved by showing the Hessian matrix of the objective function of PRIMAL (i.e., Eq. 11) is negative definite.
\end{IEEEproof}

Based on Lemma \ref{lemma2}, we use the Mixed-Integer Quadratic Programming (\emph{MIQP}) tool in the CPLEX solver to find the heuristic solution of PRIMAL.

\section{Simulation Results}
We simulate the proposed PRIMAL strategy in the cloudlet network. For comparisons, we select other two live Avatar migration decision strategies, i.e., the Follow me AvataR (\emph{FAR}) strategy and the Static strategy. The idea of the FAR strategy is to minimize the E2E delay between an Avatar and its UE by assigning the Avatar to the available cloudlet (i.e., the cloudlet has enough space to hold the Avatar), which has the lowest E2E delay to its UE's BS \cite{5}. The Static strategy is to avoid the migration cost, i.e., the locations of Avatars do not change over time after they are initially deployed. 

\begin{figure*}[!htb]
\begin{minipage}[t]{0.5\linewidth}
\centering
    \includegraphics[width=1\textwidth]{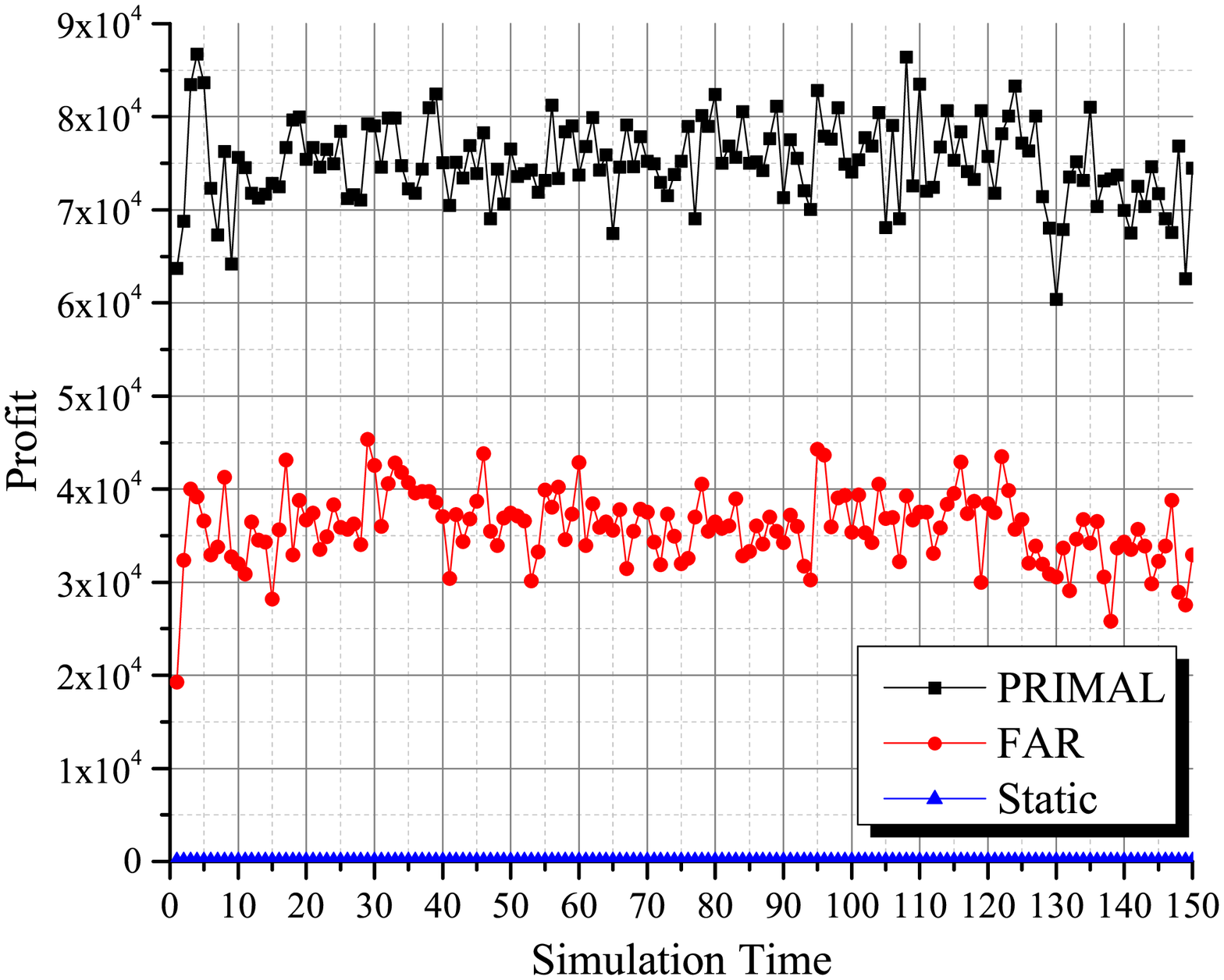}
    \caption{Profit generated by different strategies.}
    \label{fig3}
\end{minipage}
\hspace{0.1cm}
\begin{minipage}[t]{0.5\linewidth}
    \includegraphics[width=1\textwidth]{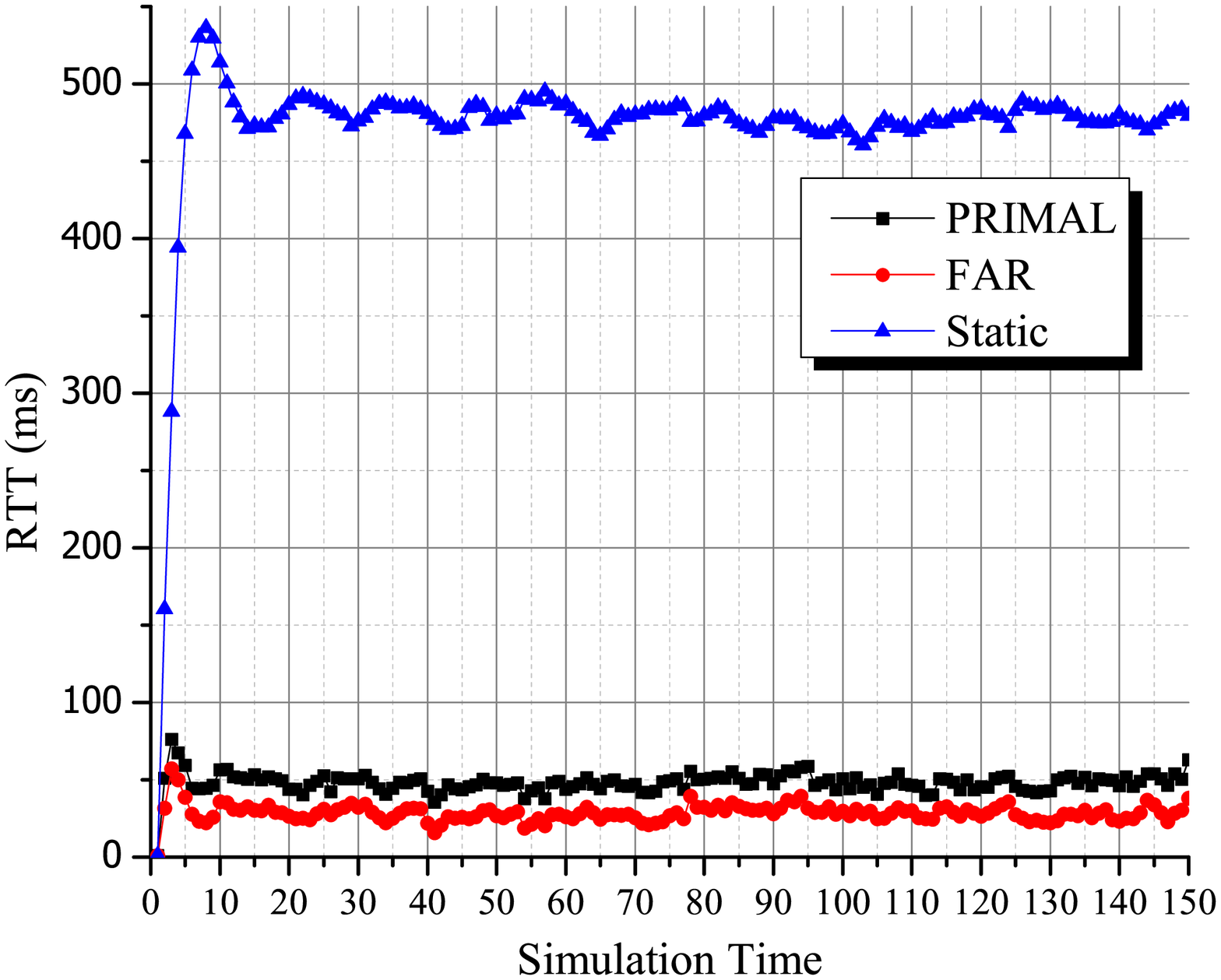}
    \caption{The average RTT between UEs and their Avatars during the simulation.}
    \label{fig4}
\end{minipage}
\end{figure*}

\begin{figure*}[!h]
\begin{minipage}[t]{0.5\linewidth}
\centering
    \includegraphics[width=1\textwidth]{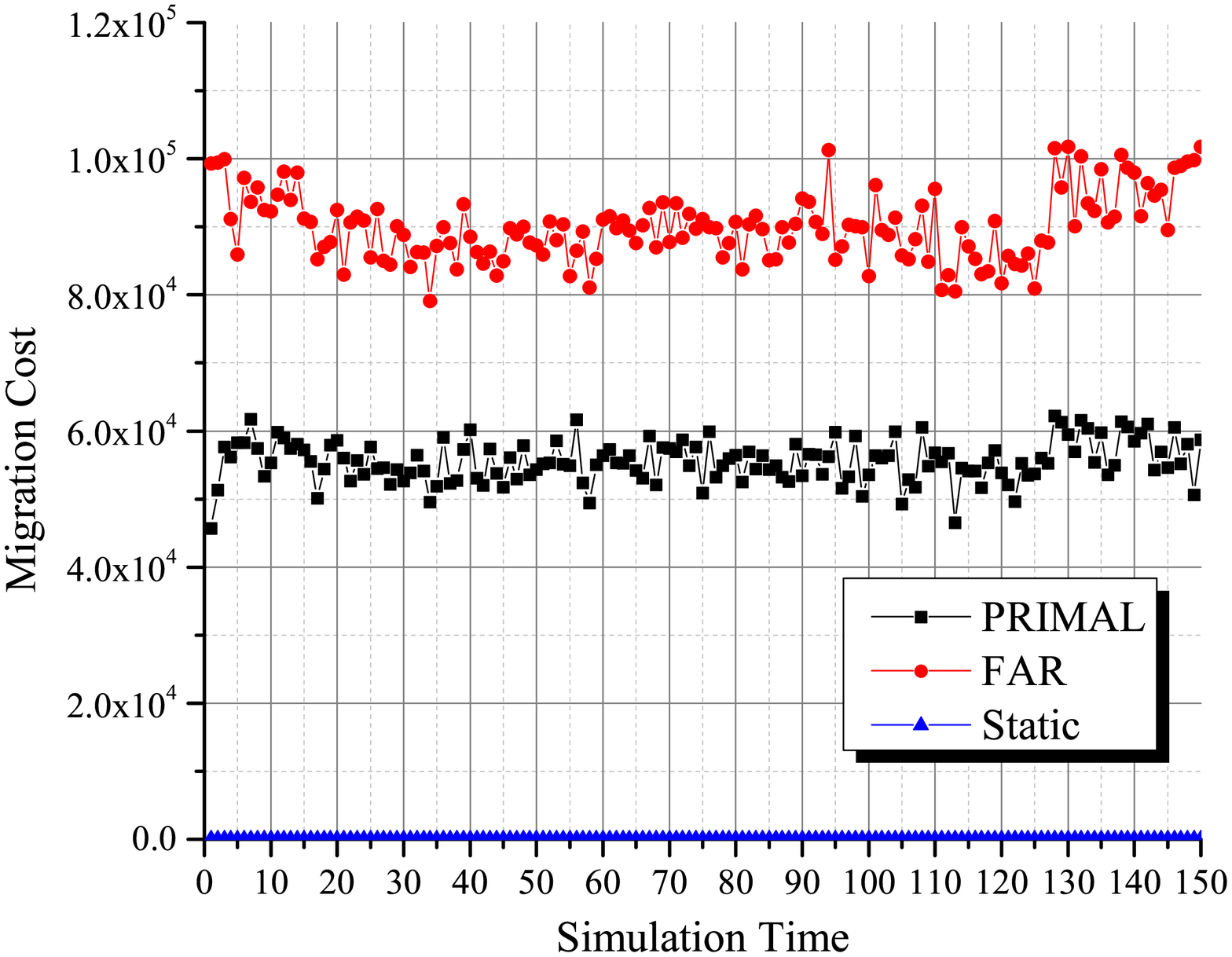}
    \caption{Migration cost incurred by different strategies.}
    \label{fig5}
\end{minipage}
\hspace{0.1cm}
\begin{minipage}[t]{0.5\linewidth}
    \includegraphics[width=1\textwidth]{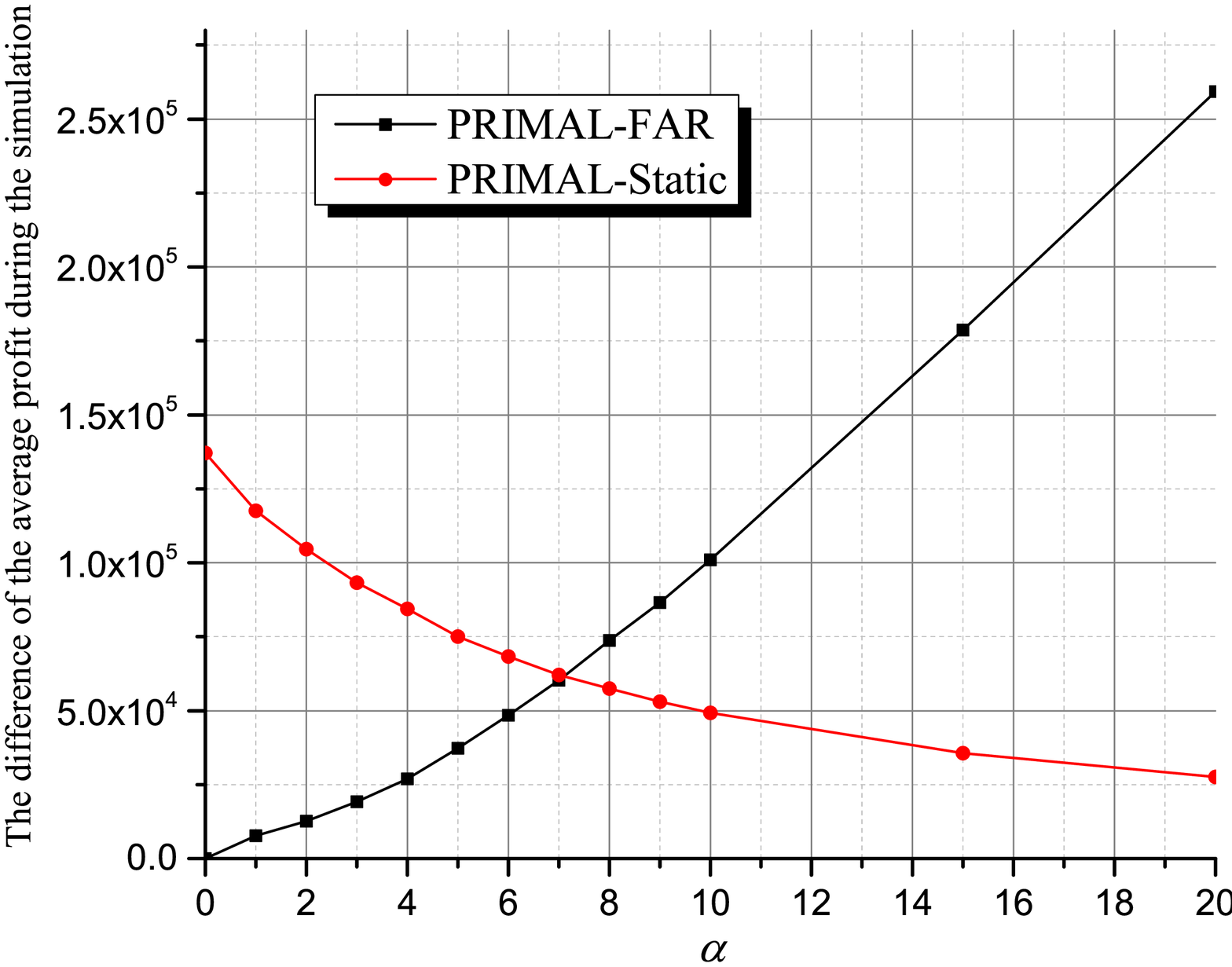}
    \caption{The average profit differences among different strategies by choosing different values of $\alpha$.}
    \label{fig6}
\end{minipage}
\end{figure*}

\begin{figure*}[!h]
\begin{minipage}[t]{0.5\linewidth}
\centering
    \includegraphics[width=1\textwidth]{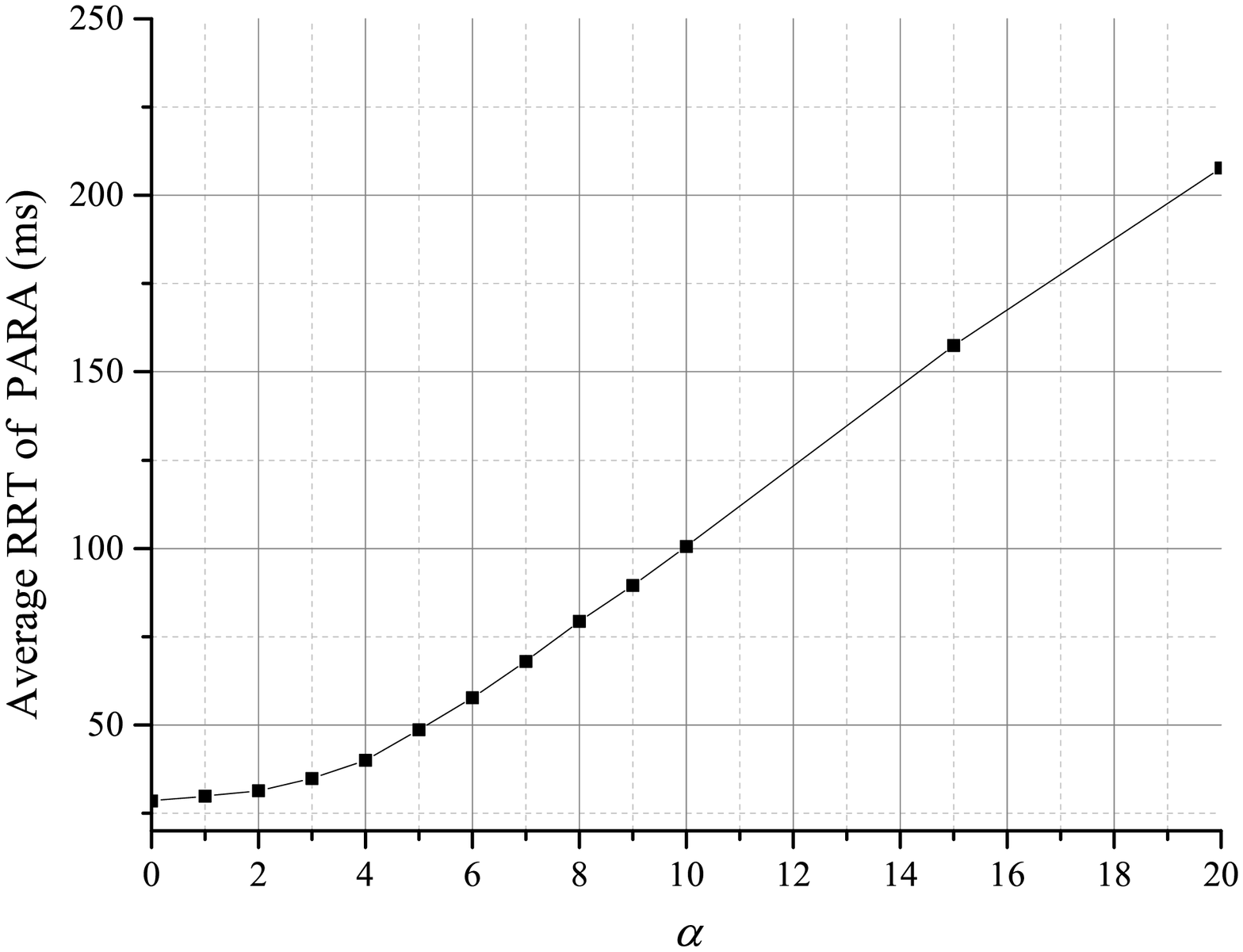}
    \caption{The average RTT incurred by PRIMAL over different values of $\alpha$.}
    \label{fig7}
\end{minipage}
\hspace{0.1cm}
\begin{minipage}[t]{0.5\linewidth}
    \includegraphics[width=1\textwidth]{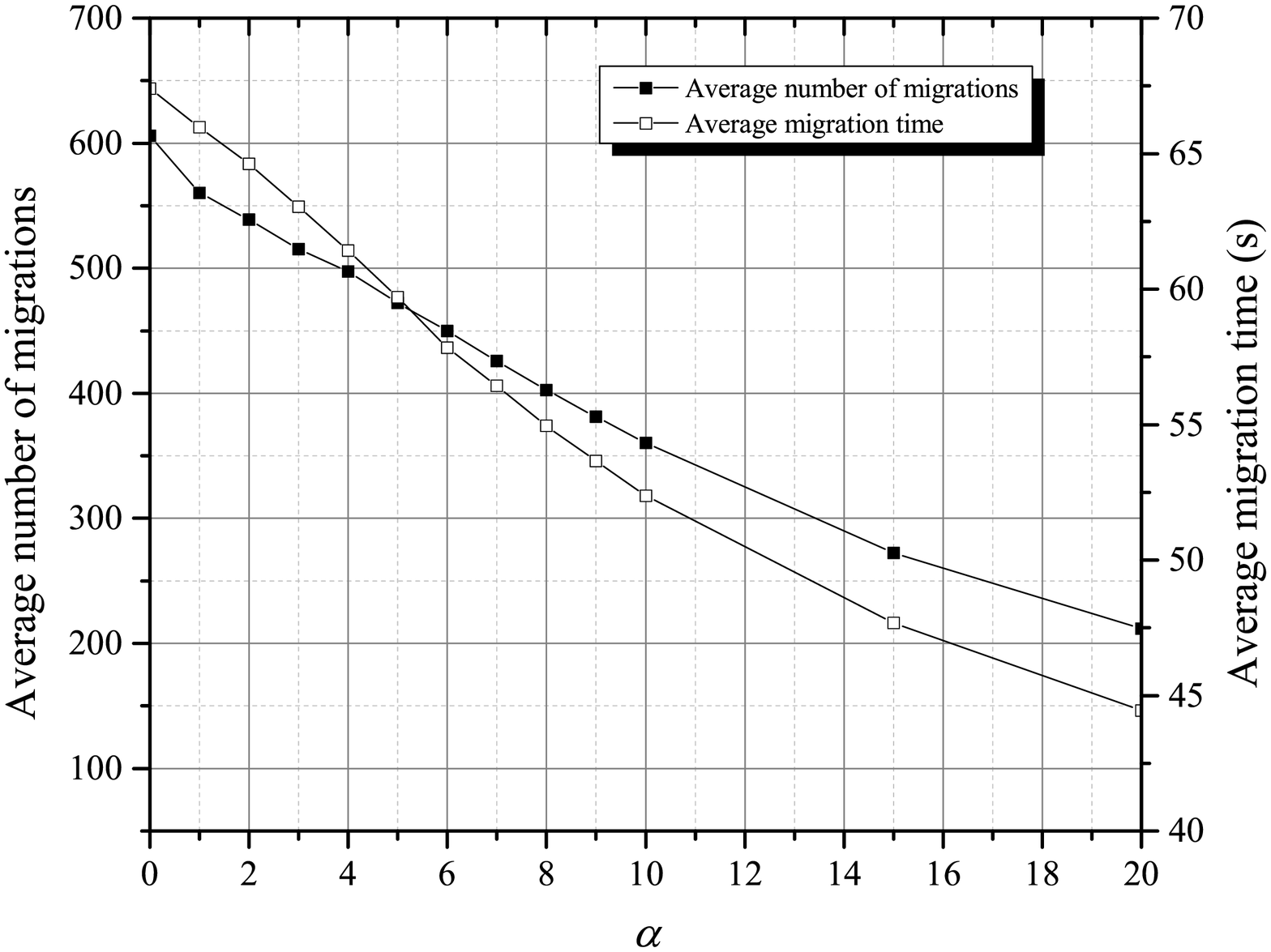}
    \caption{The average number of migrations and the average migration time incurred by PRIMAL over different values of $\alpha$.}
    \label{fig8}
\end{minipage}
\end{figure*}
    
We set up a network with the topology that includes 25 cloudlet-eNB combinations ($5 \times 5$) in a square area of 100 $km^2$. Each cloudlet connects one eNB and the coverage area of each eNB is a square area of 4 $km^2$. There are 1000 UEs, each associated with one Avatar, in the network. The UE's mobility model follows the random way point model, i.e., each UE randomly selects a destination within the network and moves toward the destination with the speed, which is randomly chosen between 0 and 10 $m/s$. Each cloudlet has the same capacity of 50 Avatars. Meanwhile, the E2E delay between a cloudlet and a BS is estimated to be proportional to the distance between them, i.e., ${d_{j,k}} = \varepsilon {\Delta _{j,k}}$ ($j \in {\bm{\mathcal{J}}}$, $k \in {\bm{\mathcal{K}}}$), where ${\Delta _{j,k}}$ is the distance between cloudlet $j$ and BS $k$, and $\varepsilon$ is the coefficient that maps the distance to the E2E delay.

The resource capacity of each Avatar is homogeneous; each Avatar is configured with 2-core CPU, 4GB memory, and 500 Mbps bandwidth. The Google cluster data trace \cite{19} is applied to emulate the CPU, memory and disk I/O utilization of each Avatar (we select the machines with CPU and memory capacity of 0.5 (normalized) in the Google cluster data trace, and calculate their CPU, memory and disk I/O utilization in each time slot. Then, the resources of Avatars (CPU, memory and disk I/O utilization time series) are emulated to be the same as those of the machines). Since the Google cluster data trace does not publish the bandwidth resource utilization and memory page dirtying rate traces of the machines, we emulate the bandwidth demand of each Avatar as a stochastic process which follows a normal distribution $N(\mu_i,\sigma_i^2)$ \cite{20}, \cite{21}, where $\mu_i$ and $\sigma_i$ are the expectation and the standard deviation of the bandwidth demand for Avatar $i$. Note that the value of $\mu_i$ and $\sigma_i^2$ are different among different Avatars, and thus we randomly select ${\mu_i} \in \left[ {0,350 \emph{Mbps}} \right]$ and $\sigma _i^2 \in \left[ {0,100 \emph{Mbps}} \right]$ for each Avatar. Furthermore, each Avatar's memory page dirtying rate depends on different types of applications running in the Avatar, i.e., some memory-intensive applications (e.g., in-memory data analytics) may generate more dirty memory pages as compared to the CPU-intensive and network I/O-intensive applications. In the simulation, the memory page dirtying rate is randomly chosen between 0 and 10K pages (each memory page consists of 32K $bits$) per time slot for each Avatar. The rest of the simulation parameters are shown in Table \ref{table1}.
\begin{table}[!htb]
	\renewcommand{\arraystretch}{1.3}
	\caption{Simulation Parameters}
	\centering
	\begin{tabular}{c c}
		\toprule
		Parameters  &  Value\\
		\midrule		
		The length of each time slot   &   5 $mins$\\
		The amount of bandwidth for doing migration, $R$ & 200 $Mbps$\\
		The migration impact factor, $w^{net}$ &  0.8 \cite{11}\\
		The migration impact factor, $w^{men}$    &  0.6 \cite{11} \\
		The migration impact factor, $w^{disk}$ &  0.4 \cite{11} \\ 
	    The migration impact factor, $w^{cpu}$       &  0.1 \cite{11}\\ 
		\bottomrule
	\end{tabular}
\label{table1}
\end{table}

Initially, each Avatar is deployed in the available cloudlet, which is the closest to its UE. First, we set up the penalty coefficient $\alpha=5$ and run the simulation. Fig. \ref{fig3} shows the profit trace by applying three different Avatar placement strategies. PRIMAL achieves the highest profit as compared to the other two strategies indicating that PRIMAL can choose the valuable migrations (i.e., $f_i>0$) to maximize the profit. In order to demonstrate the benefit for maximizing the profit, we further test the average Round Trip Time (\emph{RTT}) in terms of the E2E delay between UEs and their Avatars, as shown in Fig. \ref{fig4}, PRIMAL and FAR yield the similar average RTT, which is much lower than that of Static because Static does not dynamically adjust Avatars' placement even if the E2E delays between UEs and their Avatars are unbearable. On the other hand, as shown in Fig. \ref{fig5}, the migration cost of PRIMAL is much less than that of FAR, and thus PRIMAL achieves higher profit than FAR. We further test the average number of migrations and the average migration time as shown in Table \ref{table2}. Obviously, PRIMAL reduces the average number of the migrations as well as the average migration time, indicating that PRIMAL avoids some migrations with long migration time. In other words, PRIMAL selectively migrates Avatars that improve their RTT time (in terms of the migration gain) significantly but consume less migration time (in terms of the migration cost).

\begin{table}[!htb]
	\renewcommand{\arraystretch}{1.3}
	\caption{Simulation Results}
	\centering
	\begin{tabular}{c c c}
		\toprule
	Strategies & Average number of migrations  &  Average migration time\\
		\midrule		
		PRIMAL   &  472.2 migrations/slot & 59.7 seconds\\
		FAR    &   606.1 migrations/slot & 67.4 seconds\\ 
		\bottomrule
	\end{tabular}
\label{table2}
\end{table}

Second, we try to analyze the performance of PRIMAL by choosing different values of $\alpha$. Fig. \ref{fig6} shows the differences of the average profits between PRIMAL and FAR as well as between PRIMAL and Static during the simulation by choosing different vaules of $\alpha$. When $\alpha=0$ (i.e., there is no cost for doing live Avatar migration), PRIMAL performs exactly the same as FAR (i.e., the average profit difference is zero) as both try to minimize the RTT only. However, as the value of $\alpha$ increases, the migration cost increases and PRIMAL enables less migrations to maximize the profit, thus increasing the average profit gap between PRIMAL and FAR. On the other hand, as the value of $\alpha$ increases, the average profit gap between PRIMAL and Static is decreasing since more Avatars remain static to avoid the migration cost. We believe the performance of PRIMAL and Static is the same as $\alpha \to  + \infty$. We further test the trend of the average RTT, the average number of migrations and the average migration time for running PRIMAL as the value of $\alpha$ varies. As shown in Fig. \ref{fig7} and Fig. \ref{fig8}, when the value of $\alpha$ is small, PRIMAL triggers more migrations to improve the RTT even if the migrations consume more time. However, as $\alpha$ increases, the cost for doing migration increases and PRIMAL avoids more worthless migrations (those that improve the RTT a little bit at the expense of longer migration time). Therefore, there are tradeoffs between the average RTT and the average number of migrations, and between the average RTT and the average migration time, i.e., in order to reduce the average RTT, more migrations are triggered and more migration time is consumed. Changing the value of $\alpha$ can adjust these tradeoffs.

\section{Related Work}
Various VM placement strategies have been proposed for the resource management in data centers. Wood \emph{et al}. \cite{22} proposed \emph{Sandpiper} to detect hotspots and mitigate them by migrating the VMs to lightly loaded servers. In addition to eliminate the hotspots by VM migrations, Zhen \emph{et al}. \cite{23} tried to detect overprovisioned servers (i.e., the servers' resource utilization is lower than a predefined threshold) and migrate the VMs in the overprovisioned servers to suitable servers as much as possible so that the overprovisioned servers can be shut down to save the energy. Piao and Yan \cite{24} considered that VMs and their data may be located at different physical servers in the cloud, and thus proposed a virtual machine placement strategy to minimize the data access time, i.e., placing VMs close to their data so that the access time is minimized. Shrivastava \emph{et al}. \cite{25} proposed the VM placement strategy to place the dependent VMs (i.e., the VMs with heavy interaction among them) close to each other so that the network traffic can be reduced. Rather than only considering the gain of VM placement by migrating VMs to suitable servers, many studies argued that the cost of VM migration involved in the VM placement cannot be neglected when the resource management is applied. Verma \emph{et al}. \cite{26} proposed the VM placement strategy to minimize the total power while taking the migration cost into account. The migration cost is depicted as the throughput of the migration. Hossain \emph{et al}. \cite{27} also tried to minimize the total energy consumption by utilizing the VM placement, but they modeled the migration cost as the migration energy consumption at the destination and source servers.

Our previous work \cite{5} tried to maximize the green energy utilization of the cloudlets (each cloudlet is powered by both on-grid and green energy) in the network by utilizing the Avatar migrations to adjust the energy demands among cloudlets. In this paper, we try to optimize the Avatars' placement by maximizing the profit of Avatar migrations in terms of optimizing the tradeoff between the migration gain and the migration cost. To best of our knowledge, none of the previous work considered reducing the E2E delay between a user and its VM as the gain of VM migrations, and reducing the E2E is very important in the proposed cloudlet network (the emerging network architecture) since it can substantially facilitate communications between a UE and its Avatar in meeting the QoS of MCC applications and provisioning big mobile data analysis. 

\section{Conclusion}
In this paper, we have proposed the cloudlet network architecture to reduce the E2E delay between a UE and its Avatar so as to meet the QoS of MCC applications and to provision big mobile data analysis. However, UEs are moving in the network, and so the E2E may become worse if the UE is far away from its Avatar. In order to maintain the low E2E delay, the live Avatar migration is triggered to adjust the location of the UE's Avatar. However, the migration process consumes extra resources of the Avatar that may degrade the performance of applications running in the Avatar. Therefore, we have proposed PRIMAL to maximize the profit of the Avatar migration in terms of optimizing the tradeoff between the gain and the cost of the Avatar migration. We have also demonstrated that PRIMAL achieves highest profit as compared to the other two Avatar placement strategies, i.e., FAR (which tries to minimize the E2E delay by neglecting the migration cost) and Static (which tries to minimize the migration cost without considering the E2E delay).

\bibliographystyle{IEEETran}

\end{document}